\documentclass[12pt]{article}
\usepackage{amsmath}
\usepackage{graphicx}
\usepackage{enumerate}
\usepackage{natbib}
\usepackage{url} % not crucial - just used below for the URL 

\usepackage[figuresright]{rotating}
\usepackage[english]{babel}
\usepackage[utf8x]{inputenc}
\usepackage[T1]{fontenc}
\usepackage{enumitem}
\usepackage{fancyhdr}
\usepackage{adjustbox,lipsum}
\usepackage{amsmath,amssymb,bm}
\usepackage{etoolbox}
\usepackage{float}
\usepackage{color}
\usepackage{accents}
\usepackage{algorithm,algpseudocode}
\usepackage{graphicx}
\usepackage{tabularx}

\newtheorem{theorem}{Theorem}
\newtheorem{corollary}{Corollary}
\algnewcommand{\To}{\textbf{To }}
\algnewcommand\Input{\item[\textbf{Input:}]}
\algnewcommand\Initialize{\item[\textbf{Initialize:}]}

\def\bSig\mathbf{\Sigma}

\newcommand{\T}{\intercal}

\newcommand{\blind}{1}
\usepackage{fancyhdr}
\usepackage{lipsum}% just to generate text for the example

\fancypagestyle{specialfooter}{%
	\fancyhf{}
	
	\fancyfoot[L]{CONTACT: Min Qian, PhD, mq2158@cumc.columbia.edu, Department of Biostatistics, Columbia University, 722 w. 168th St, Fl 6, New York City, NY, 10032, USA. The authors gratefully acknowledge \textit{NIH grants  R01MH109496 and R21MH108999}.}
}

\begin{document}

	\def\spacingset#1{\renewcommand{\baselinestretch}%
		{#1}\small\normalsize} \spacingset{1}

	%%%%%%%%%%%%%%%%%%%%%%%%%%%%%%%%%%%%%%%%%%%%%%%%%%%%%%%%%%%%%%%%%%%%%%%%%%%%%%
	
	\if1\blind
	{
		\title{\bf Personalized Policy Learning using Longitudinal Mobile Health Data}
		\author{Xinyu Hu, Min Qian, Bin Cheng and Ying Kuen Cheung
			%\thanks{The authors gratefully acknowledge \textit{NIH grants  R01MH109496 and R21MH108999}.}
			\hspace{.2cm}\\
			Department of Biostatistics, Columbia University}
		\date{}
		\maketitle
	} \fi
	
	\if0\blind
	{
		\bigskip
		\bigskip
		\bigskip
		\begin{center}
			{\LARGE\bf Personalized Policy Learning using Longitudinal Mobile Health Data}
		\end{center}
		\medskip
	} \fi
	
	\bigskip
	\begin{abstract}
		We address the personalized policy learning problem using longitudinal mobile health application usage data.  Personalized policy represents a paradigm shift from developing a single policy that may prescribe personalized decisions by tailoring.  Specifically, we aim to develop the best policy, one per user, based on estimating random effects under generalized linear mixed model.  With many random effects, we consider new estimation method and penalized objective to circumvent high-dimension integrals for marginal likelihood approximation.  We establish consistency and optimality of our method with endogenous app usage.  We apply our method to develop personalized push (``prompt'') schedules in 294 app users, with a goal to maximize the prompt response rate given past app usage and other contextual factors.  We found the best push schedule given the same covariates varied among the users, thus calling for personalized policies.  Using the estimated personalized policies would have achieved a mean prompt response rate of 23\% in these users at 16 weeks or later: this is a remarkable improvement on the observed rate (11\%), while the literature suggests 3\%-15\% user engagement at 3 months after download.  The proposed method compares favorably to existing estimation methods including using the R function ``glmer'' in a simulation study.
	\end{abstract}
	
	\noindent%
	\vfill

\thispagestyle{specialfooter}

	\newpage
	\spacingset{1.5} % DON'T change the spacing!

\section{Introduction} \label{sec:intro}
%	\mincomment{Ken, one thing to note that when I edit the manuscript,  I used "treatment", "action", "decision" indifferently as I'm not sure which one is the best. Conventionally, we use "treatment", but the decision in our example is not treatment. You can help to decide what to use in the paper.}
Mobile technologies such as smartphones and wearables enable continuous monitoring of 
exposure to environmental stressors and ecological assessment of health-relevant data over an extended period of time,  thereby facilitating the delivery of  tailored intervention in an adaptive manner \citep{riley2011health}. 
Examples  abound.
%\citet{piwek2016rise} suggest giving patients direct access to the personal health analytics so as to aid in preventive care and disease management.
\citet{heron2010ecological} review the use of tailored interventions based on momentary assessments
 to support management 
of  a variety of  health behaviors and symptoms such as smoking,  diabetes,  and weight loss.
\citet{depp2010mobile} study the efficacy of personalized pushed engagement based on real-time data in mental illness patients.
\citet{mohr2013continuous} envision a continuous evaluation system of health apps 
based on evidence generated by routinely collected data.  To illustrate, %these concepts, % concrete terms,
we consider  a suite of smartphone apps (called IntelliCare) that serves users with anxiety or
depression using different psychological treatment strategies 
including cognitive behavioral therapy, positive psychology, and physical activity-based interventions  \citep{mohr2017intellicare}.
The suite consists of a Hub app that helps users navigate apps within the IntelliCare ecosystem and coordinate their experience, with a specific function to provide links and recommendations for other IntelliCare apps so as to maximize user engagement based on a user's app usage history \citep{cheung2018}.
In this article, we are motivated by a sub-study of the IntelliCare suite, 
in which the Hub app would send pushed notifications to prompt a user to complete a short four-item patient health questionnaire 
repeatedly on 7-day intervals at a random time during the day.
While the purpose of the prompts is to remind user to assess their  depression and anxiety symptoms,
the response rate  was expected to be modest and declining quickly over time based user engagement
reported  in the literature  \citep{cgf2009, helander2014}.  
Since time of day is a known  factor of mobile application usage \citep{angrybird}, 
the objective of this study is to learn the best time period 
to push the prompt (policy) that maximizes response given other contextual factors
a user experiences as well as the user's past engagement.
%ne of this study's aim is 
%to learn the best time to push the reminder so as to maximize response given other contextual factors experienced  by a user as well
%as the user's past engagement. 
In addition, since there is often unobserved 
between-user heterogeneity due to a user's own circumstances that is difficult to capture or measure 
 \citep{heter2017}, the eventual goal is to develop policies, one for each user, that can provide 
 personalized feedback through their interaction with the IntelliCare apps.

Numerous policy learning methods  that support  decision making using medical data and mobile health  data have been proposed. 
For example,  there is a large statistical literature  on  reinforcement learning algorithms that 
estimate optimal policies under a {nomothetic} model \citep{murphy2003optimal, qian2011performance, zhang2012robust,laber2014iq, song2014penalized,zhao2012estimating, zhao2015new, ertefaie2018constructing, luckett2019estimating}. 
A nomothetic approach assumes that a population model captures all between-subject heterogeneity and facilitates estimation
by pooling data across participants.  While this approach may  address user heterogeneity and allow for the estimation of personalized policies by incorporating appropriate interactions with the actions, it often requires 
the untestable assumption of no unobserved confounders.
Alternatively, an ideographic approach achieves personalization using an ``N-of-1'' approach
whereby a person's own decision model is estimated using the person's own data only
  \citep{nof1-1, duan2014,lei2014actor}. 
Although this approach in principle allows for insights about individuals without assumptions about any reference population,
its practicality relies on how long a user can be followed.  In general, the efficiency of this approach may suffer,
especially  in situations where an action exhibits similar effects on all individuals.

%We are interested in the effect of current action on the immediate outcome. In mHealth area, such as Intellicare study, a prompt message asking for filling a questionnaire requires an immediate response, as it assesses a user's current psychological condition. However, the expected outcome may depend on potential time varying confounders which might be influenced by past actions. The causal effect of current action relates to treatment blip defined in the structural nested mean model \citep{robins1989analysis}. The effect is conditional on all historical information and with respect to a prespecified future with reference actions that define the distribution for all future actions. In our setting, the history affects the current outcome through the current potential confounders of interest. Thus the estimation of the current outcome conditions on the confounders. As in Intellicare study the prompt is sent randomly and independently of historical information, we specify that the current action does not affect the distribution of future actions.

In this article, we consider estimating personalized policies under the generalized linear mixed model (GLMM) framework with the
 outcome at each time point as the dependent variable and time-varying covariates, action and their interactions as the
predictors.  For instance, in the  IntelliCare ``Prompt'' sub-study, the outcome of interest is a binary response and the action
is the time period during a day when a prompt is pushed.
The estimated policy aims to recommend an action that maximizes the predicted outcome  based on 
the contextual factors experienced by  a user and the user's past engagement.
In addition to tailoring, each user will have a personalized policy through the estimation of the random effects, which capture individual
departure from the population model due to unobserved heterogeneity.

%\mincomment{I suggest not to use the phrase "An exception" in line 8 of the paragraph below. There are some other literature on this issue. And they proposed different conditions. I didn't mention those here because those literature didn't study treatment effects.  For example, in the linear model setting, Pepe and Anderson (1994) showed that the GEE estimate has conditional interpretation if a working diagonal covariance structure is used, or if $E[S(Y_{it}|\underline S_{iT}]=E[Y_{it}|S_{it}]$ (i.e. conditional mean of Y given past and future states is the same as conditional mean given present state).}

While GLMM is one of the most popular methods to handle longitudinal outcome data,  GLMM-based estimation methods are 
largely designed for settings where the covariates are exogenous with respect to the outcome process.  
When the time-varying covariates are allowed to be endogenous, that is, letting them depend on the outcome process, previous
treatment assignments, and possibly random effect parameters, estimation  of the GLMM fixed effect coefficients---based on likelihood or
generalized estimating equations---may lead to bias, because it no longer corresponds to the 
conditional interpretation of the parameters % in the GLMM may no longer hold under such endogeneity
see \citet{sullivan1994cautionary} and \citet{diggle2002analysis} for example. 
In the case of linear mixed models, when
%An exception is  \citet{qian2019linear} who study 
the conditional interpretation of fixed effects is consistent with the scientific interest in predicting person-specific effects,
\citet{qian2019linear}
show that standard software %for linear mixed models can 
can be used to obtain a valid estimate of the fixed effects
if the time-varying covariates are independent of the random effects parameters conditional on past history. 
In this article, we examine the conditions under which the proposed estimation method 
 work in the presence of endogeneity in GLMM.   Furthermore, as it will be shown in Section~\ref{sec:thm}, our method 
does not require  a full conditional distribution of outcome or random effects to be correctly specified, but relies on a much weaker
assumption that the conditional mean outcome model is correctly specified.

We note some previous work on estimating personalized treatment using GLMM. For example, \citet{cho2017personalize} use GLMM to predict individual outcome under each treatment arm with a random slope on the treatment indicator, and build a random forest model to predict random slope using patients baseline covariates. Personalized treatment can then be implemented by selecting the treatment with the maximal estimated random effects.
However, little if any of the previous work includes random effects for treatment-by-covariate interactions in the model, thus having no
provision for tailoring.  Allowing for random effects for treatment-by-covariate interactions  presents a key computational
challenge, as most methods rely on approximating of the marginal likelihood of the outcomes by integrating out  the random effects.
  When there are moderate or large number of random effects terms, standard GLMM software fail to produce accurate approximation of the integrals. To address the computational challenge, we propose a novel algorithm that 
  estimates the fixed effects and random effects  jointly with a ridge-type penalty on the latter.
    In addition, to avoid overfitting
  individual deviations from the population mean, we propose to apply a group lasso penalty 
 on the random effects     \citep{yuan2006model}.  
  This
  penalized  approach is critical in circumventing the large number of random effects for treatment-by-covariate interactions.

This article is organized as follows. In Section \ref{sec:prelim}, we set up the formulation of the personalized policy learning problem, and present new policy estimation methods.
We then study the theoretical properties of the proposed method in Section \ref{sec:thm}, and compare it with some existing approaches in Section \ref{sec:simu}.
We will revisit the IntelliCare Prompt study in Section \ref{sec:intell} and apply the proposed method to develop personalized policies in the study.
We end this article with some concluding remarks in Section~\ref{sec:diss}.
Details of computational algorithms, technical derivations, and proofs are provided as separate Supplementary Materials.

 %personalized policy learning (PPL) framework, present the estimation procedure and the learning algorithm, and make connection to standard GLMM literature. In Section \ref{sec:thm}, we study the asymptotic behavior of the estimated parameters and policies. In Section \ref{sec:exp}, we conduct extensive
%simulation studies to evaluate the performance of PPL by comparing it with other competing methods, and applied those methods on Intellicare data to estimate the right time for sending prompts to users for improving the response rate. We conclude the paper with discussion and future work in Section \ref{sec:diss}. All proofs and extra simulation results are presented in the Appendix.

%%%%%%%%%%%%%%%%%%%%%%%%%%%%%%%%%%%%%%%%%%%%%%%%%%%%%%%%%%%%%%%
\section{Personalized Policy Learning} \label{sec:prelim}
    \subsection{Notations and Problem Formulation} \label{ssec:framwork}

Suppose mobile application user $i$ is tracked longitudinally over  $m_i$ time points.  At time  $t$,
an action $A_{it}$ taking values in a pre-specified finite discrete action space $\mathcal{A}$ is randomized to the user, with
a vector of covariates $S_{it}\in\mathcal{S}$ observed prior to the action.
Let $Y_{it}$ denote the outcome of interest observed after each action, with the convention that large values of $Y_{it}$ are good.
We note that the covariates $S_{it}$ may include endogenous variables that depend on previous outcomes and actions, 
as well as other exogenous and contextual factors.
In summary, the trajectory of each user is denoted by the triplets  $\left\{ (S_{it}, A_{it}, Y_{it}) : t=1, \ldots, m_i \right\}$.
We further denote the entire history up to $t$ by $\underline S_{it} = (S_{i1},\ldots, S_{it})$ and  $\underline A_{it} = (A_{i1},\ldots, A_{it})$.

Our objective is to estimate for a given user $i$ a personalized policy $\pi_{0i}$, which when implemented will result in the maximal conditional expected outcome, $E_{\pi_{0i}} (Y_{it}|\underline S_{it}, \underline A_{i,t-1})$, 
where the expectation is taken with respect to the conditional distribution of $Y_{it}$ given the  history $(\underline S_{it}, \underline A_{i,t-1})$ and action $A_{i,t}$ is consistent with $\pi_{0i}$.  
% We are making myopic decision. That is, at current decision point $t$, given current information $(\underline S_{it}, \underline A_{i,t-1})$, we choose $A_{it}=\pi_{0i}$ so as to maximize current expected outcome $Y_{it}$. So $\pi_{0i}$ would be a function mapping from $(\underline S_{it}, \underline A_{i,t-1})$ to $A_{it}$, and its form may vary across $t$. Under Markovian assumption, $\pi_{0i}$ is a mapping from $S_{it}$ to $A_{it}$. Further under stationary condition, we say its form is time-invariant, and we can use $\pi_{0i}$ instead of $\pi_{0i,t}$.
We  further make the commonly used assumption that the conditional distribution of $Y_{it}$ given 
 $\underline S_{it}, \underline A_{it}$ is Markovian, so that 
 $E_{\pi_{0i}} (Y_{it}|\underline S_{it}, \underline A_{i,t-1})  = E_{\pi_{0i}}  (Y_{it}|S_{it}) = E \{ Y_{it}|S_{it}, A_{it} = \pi_{0i}(S_{it}) \}$.
%$E_{\pi_{0i}} (Y_{it}|\underline S_{it}, \underline A_{i,t-1})  = E_{\pi_{0i}}  (Y_{it}|S_{it}) = E (Y_{it}|S_{it}, A_{it} = \pi_{0i}(S_{it}))$. 
%The first equality is due to Markovian, and the second equality is based on the set of causal assumptions, which are valid if $A$ is randomized.
 %$E_{\pi_{0i}} (Y_{it}|\underline S_{it}, \underline A_{it}) = E (Y_{it}|S_{it}, A_{it} = \pi_{0i}(S_{it}))$.
 We note that $S_{it}$ can include lagged variables at previous time points (e.g., $Y_{i, t-2}$).
 Further let 
$Q_{0i}(s,a) = E(Y_{it}|S_{it}=s, A_{it}=a)$ so as to make explicit  the conditional
expectation is user-specific.
Then $\pi_{0i}(s) \in\arg\max_{a\in\mathcal{A}}Q_{0i} (s,a)$.
Once  $\pi_{0i}$ is estimated by $\hat \pi_i$ (say), the estimated policy will be used to guide decision making for the user in the future time points.
While this formulation of the problem assumes a stationary policy in that the function $Q_{0i}$ is time-invariant, 
the policy decisions can be time-dependent by including time in the covariate state $S_{it}$.  In our application, this assumption is   aligned  with the fact that mobile application usage is  habitual given other contextual factors.  %\citep{angrybird}.

%To operationalize the policy decisions, we postulate 
We  facilitate the learning problem under the GLMM  framework, and postulate 
\begin{align}
Q_{0i}(S_{it}, A_{it})  = &\,  g^{-1} \left\{ h_1(S_{it}, A_{it})^\T\bm{\beta}_0 + h_2(S_{it}, A_{it})^\T\bm{\alpha}_{0i\cdot}^\T \right\}
:= Q(S_{it}, A_{it}; \bm{\beta}_0, \bm{\alpha}_{0i\cdot}),   \label{eqn:mixed1}
\end{align}
for $i=1,\ldots, n$ and $t=1,\ldots, m_i$,
where $g(\cdot)$ is a known strictly monotone increasing link function. For example, the canonical forms of $g(\cdot)$ 
are respectively the identity function for continuous outcome,  logit for binary outcome, and logarithmic for counts. 
Here
$\bm{\beta}_0$ is a $p$-dimensional vector of unknown parameters, and $h_1(S_{it}, A_{it})\in\mathbb{R}^p$ is  a pre-specified vector function of $(S_{it}, A_{it})$ so that $h_1(S_{it}, A_{it})^\T\bm{\beta}_0$ is the fixed effects component; for example, 
$h_1(S_{it}, A_{it})^\T\bm{\beta} = \beta_0 + \beta_1S_{it} + \beta_2A_{it} + \beta_3S_{it}A_{it}$. The random effects
are denoted by $\bm\alpha_0$, an $n\times q$ ($q\leq p$) matrix with the $i$-th row, $\bm\alpha_{0i\cdot}$, denoting the random effects parameters for the $i$-th user, and $h_2(S_{it}, A_{it})\in\mathbb{R}^q$ is a sub-vector of $h_1(S_{it}, A_{it})$ chosen so that $h_2(S_{it}, A_{it})^\T \bm\alpha_{0i\cdot}$
models subject-specific deviations from the mean model. Under model (\ref{eqn:mixed1}) and a monotone increasing $g(\cdot)$, the optimal policy $\pi_{0i}$  can  be expressed as
\begin{equation}
\pi_{0i}(S_{it})\in\arg\max_{a\in\mathcal{A}} \left(h_1(S_{it}, a)^\T\bm\beta_0 + h_2(S_{it}, a)^\T \bm\alpha_{0i\cdot}^\T\right).
\label{eqn:pi_star}
\end{equation}

Note that $\bm\alpha_{0i\cdot}$ play dual roles in our proposed method. On one hand, it defines the individual deviation from the mean model 
of the $i$-th user, and can be viewed as a fixed parameter to be estimated and to be acted upon. 
This role operationalizes the personalized policy decisions (\ref{eqn:pi_star}).
%Thus, when a user is specified, $\bm\alpha_{0i\cdot}$ is viewed as a fixed parameter. 
 On the other hand, %since users $i$ for $i=1,\ldots, n$, are randomly selected from the population,
  $\{ \bm\alpha_{0i\cdot} \}$ can  be viewed as a random sample of the population.  This viewpoint motivates
some degree of ``smoothness'' in  the estimation of $\bm\alpha_{0i\cdot}$'s, which is described next.

%independent and identically distributed (i.i.d.) latent random vectors from the population point of view.

%
%To enhance the robustness of the model, the actions are centered by subtracting the propensity score from the indicator of the action, i.e. $\mathbb{I}(A_{it}=k) - p(k|H_{it})$ with $K$ actions $A_{it} \in \{1,\dots,K\}$ and $K-1$ centered terms $k \in 1,\dots, K-1$. The centering yields orthogonality between the estimated parameter in the action effect and the conditional mean function. That is, if the conditional mean function is misspecified, the estimated action effect will still be consistent.

\subsection{Policy Estimation} \label{ssec:policy}

%We consider modeling $E(Y_{it}|S_{it},A_{it})\triangleq Q_{0i}(S_{it}, A_{it})$ by
%\begin{align}
%Q(S_{it}, A_{it}; \bm{\beta}, \bm{\alpha}_{i\cdot})  = &\,  g^{-1}[h_1(S_{it}, A_{it})^\T\bm{\beta} + h_2(S_{it}, A_{it})^\T\bm{\alpha}_{i\cdot}^\T],  \label{eqn:model}
%\end{align}
%where $\bm{\beta}\in\mathbb{R}^p, \bm{\alpha}_{i\cdot}^\T\in\mathbb{R}^q$.
%\mincomment{In the first sentence below, Bin and I suggest to add "working" before "conditional log-likelihood". So it reads like "Let ... denote the working conditional log-likelihood of...".  In this way, people kind of know that it is a working model. Otherwise, as of current, the first impression is that $\ell$ need to be the true log-likelihood, although later we said it is not necessary.}

Let $\{\ell(Y_{it}, S_{it}, A_{it}; \bm{\beta}, \bm{\alpha}_{i\cdot}, \phi): \bm{\beta}\in\mathbb{R}^p, \bm{\alpha}_{i\cdot}^\T\in\mathbb{R}^q\}$ denote the working conditional log-likelihood of $Y_{it}$ under a fully specified GLMM with the systematic component (\ref{eqn:mixed1}).
For example, with a continuous  $Y_{it}$, we may set $\ell(\cdot)$ to be the Gaussian log-likelihood  with mean 
$Q(S_{it}, A_{it}; \bm{\beta}, \bm{\alpha}_{i\cdot})$,  variance $\sigma^2$, and an identity link.
When $Y_{it}$ is binary, we may choose 
$\ell(\cdot)$ to be the Bernoulli log-likelihood with probability $Q(S_{it}, A_{it}; \bm{\beta}, \bm{\alpha}_{i\cdot})$ and an logit link.
However, the theoretical results described in Section~\ref{sec:thm} will hold for any choice of $\ell(\cdot)$ that 
satisfies
\begin{align}
& E\Big[\sum_{i=1}^n\sum_{t=1}^{m_i}\nabla_{\bm{\beta}}\ell(Y_{it}, S_{it}, A_{it}; \bm{\beta}, \bm{\alpha}_{i\cdot}, \phi)\big|_{\bm{\beta}=\bm{\beta}_0, \bm{\alpha}=\bm{\alpha}_{0}}\Big]=\bm{0} \nonumber\\
\mbox{ and } & E\Big[\sum_{t=1}^{m_i}\nabla_{\bm{\alpha}_{i\cdot}}\ell(Y_{it}, S_{it}, A_{it}; \bm{\beta}, \bm{\alpha}_{i\cdot}, \phi)\big|_{\bm{\beta}=\bm{\beta}_0, \bm{\alpha}=\bm{\alpha}_{0}}\Big]=\bm{0}  \mbox{ for } i=1,\ldots, n,
\label{eqn:focondition}
\end{align}
where $\bm{\alpha}_{i\cdot}$ is the $i$-th row of $\bm{\alpha}$,  $\phi$ is a nuisance parameter in the working log-likelihood, and $\nabla_{\bm{\beta}}\ell$ and $\nabla_{\bm{\alpha}_{i\cdot}}\ell$ denote the partial derivatives of $\ell$ with respect to $\bm{\beta}$ and $\bm{\alpha}_{i\cdot}$, respectively.
It is easy to verify that the Gaussian and the Bernoulli log-likelihoods satisfy (\ref{eqn:focondition}); and since they are often the 
practical choices for continuous and binary outcomes,  they may be used as pseudo-log-likelihood in many applications.
 Correspondingly, we define the penalized pseudo-log-likelihood
\begin{align}
L_{ppl}(\bm{\beta}, \bm{\alpha}) =  \sum_{i=1}^n\sum_{t=1}^{m_i}\ell(Y_{it}, S_{it}, A_{it}; \bm{\beta}, \bm{\alpha}_{i\cdot},\phi) - \frac{1}{2}\sum_{i=1}^n\bm{\alpha}_{i\cdot}\bm{D}^{-}\bm{\alpha}_{i\cdot}^\T -\lambda \sum_{l=1}^qw_l\|\bm{\alpha}_{\cdot l}\|,
\label{eqn:ppl}
\end{align}
where $\bm{D}\in\mathbb{R}^{q\times q}$ is a symmetric positive semi-definite matrix, $\bm{D}^{-}$ is the Moore-Penrose generalized inverse of $\bm{D}$, and $\lambda\geq 0$ is a tuning parameter. % which is selected to minimize an AIC-type information criterion.

We propose to estimate $\bm\beta_0$ and $\bm\alpha_0$ by maximizing (\ref{eqn:ppl}). The maximum penalized-pseudo-likelihood estimator
is denoted by
\begin{align}
(\hat{\bm{\beta}},\hat{\bm{\alpha}}) = \arg\max_{\bm{\beta}\in\mathbb{R}^p, \bm{\alpha}\in\mathbb{R}^{n\times q}} L_{ppl}(\bm{\beta}. \bm{\alpha}),
\label{eqn:apql}
\end{align}
and  the corresponding  personalized policy for user $i$ is estimated by
$$\hat\pi_i(s)\in\arg\max_{a\in\mathcal{A}}  \left(h_1(s, a)^\T\hat{\bm\beta} + h_2(s, a)^\T \hat{\bm\alpha}_{i\cdot}^\T\right),$$
analogously to $\pi_{0i}$ in (\ref{eqn:pi_star}).

The second term on the right hand side of (\ref{eqn:ppl}) puts a ridge-type penalty   to shrink
and stabilize the estimation of the random effects $\bm{\alpha}\in\mathbb{R}^{n\times q}$.  Under the 
viewpoint that  $\{ \bm{\alpha} \}$ is a random sample of a population, it is natural to choose
 $\bm{D}$ to reflect the variance-covariance matrix of $\bm{\alpha}_{0i\cdot}^\T$, although it is not required 
 for the asymptotic properties to hold (see Section~\ref{sec:thm}).
 The third term in (\ref{eqn:ppl})  is the group lasso penalty, where each group $l$ contains the random effects parameter of the $l$-th term in $h_2(S_{it}, A_{it})$ for all $n$ users.  Under a similar viewpoint, it is intuitive to set the group-specific weight $w_l\geq 0$
 to be inverse proportional to the variance of $\bm{\alpha}_{il}$. 

In practice, 
%As conventionally in mixed effects models, here we assumed that $\bm{D}$, $\phi$ and $w_l$'s are given. 
%In practice, 
we propose to update $\bm{D}$, $\phi$ and $w_l$'s iteratively, in conjunction with the trust region newton 
(TRON) algorithm in  the estimation of $\bm{\beta}$ and $\bm{\alpha}$.  Briefly, the TRON algorithm combines the trust region method
\citep{steihaug1983conjugate} and the truncated newton method \citep{nash2000survey} to solve an unconstrained
convex optimization problem.  At each iteration,  TRON defines a trust region and 
approximates the objective function using a quadratic model within the region. If a pre-specified change of the objective function
is achieved in the current iteration, the updated direction is accepted and the region is expanded; the region will be shrunk otherwise. 
The approximation sub-problem is solved via the conjugate gradient method. 
Since TRON solves the inverse of a potentially large Hessian matrix by iteratively updating the parameters, 
convergence can be achieved quickly with a large and dense  Hessian.  Overall, the computational cost per iteration is of
the order of the number of nonzero elements in the design matrix.
In addition, we propose to choose the  tuning parameter $\lambda$ for the group lasso penalty using an AIC-type criterion.
The details are given in Section S1 of the Supplementary Material.

\section{Theoretical Remarks}
%\subsection{Asymptotic Properties}
\label{sec:thm}
In this section, we  study the asymptotic behavior of $\hat{\bm{\beta}}$ and $\hat{\bm{\alpha}}$, and %. And then, we present the 
conditional and marginal performance of estimated policies $\hat{\pi}_i$'s under the following  assumptions. All proofs are given in Section S2 of the Supplementary Material.
\begin{enumerate}[label=(C\arabic*)]
	%\item The maximizer of $E\left \{ \ell_{{\rm APQL}}(\bm{\beta}, \bm{\alpha})|\bm{\alpha}\right \}$, $(\bm{\beta}_0, \bm{\alpha}_0)$, is an interior point of a compact set $\Omega \in \mathbb{R}^{p + q}$.
	\item There exists a positive constant $c_1$, such that  the treatment randomization probability $P(A_{it}=a_t|\underline S_{it}=\underline s_t, \underline A_{i,t-1}=\underline a_{t-1}) \geq c_1$  for all possible values of $(\underline s_t, \underline a_t)$ at any time point $t\geq 1$.
	\label{ap:treatment}
	
	\item The random vectors $h_1(S_{it},A_{it})$ and $h_2(S_{it},A_{it})$ and outcome $Y_{it}$ are square integrable under the data generative distribution for $t\geq 1$ and $i=1, \ldots, n$.
	\label{ap:integrable}
	
	\item The latent random effects $\bm{\alpha}_{0i\cdot}, i=1, \ldots, n$, are independent and identically distributed with mean $\bm{0}$ and finite variance $\bm{\Sigma}$.
	\label{ap:alpha}
	
	\item There exists $(\bm{\beta}_0, \bm{\alpha}_0)$ such that (\ref{eqn:mixed1}) holds, and $(\bm{\beta}_0, \bm{\alpha}_0)$ is $P_{\bm{\alpha}_0}$-almost surely an interior point of a compact set $\Omega \in \mathbb{R}^{p + nq}$.
	\label{ap:model}
	
	\item The pseudo-log-likelihood $\ell(Y_{it}, S_{it}, A_{it}; \bm{\beta}, \bm{\alpha}_{i\cdot},\phi)$ is concave in $(\bm{\beta},\bm{\alpha})$, satisfies condition (\ref{eqn:focondition}), and  its expected second order derivative is continuous in $(\bm{\beta},\bm{\alpha})$.
	\label{ap:ppl}
	%The working model of $Y_{it}$ at $(\bm{\beta}_0, \bm{\alpha}_0)$, that is, $g\{ E(Y_{it}|S_{it}, A_{it}; \bm{\beta}_0, \bm{\alpha}_{0i\cdot})\}  =  h_{1}(S_{it}, A_{it})^\T\bm{\beta}_0 + h_{2}(S_{it}, A_{it})^\T\bm{\alpha}_{0i\cdot}^\T$ implies
	%     $ E\left \{  \nabla \ell_1 (\bm{\beta}_0,\bm{\alpha}_0)\right \} =\bm{0}$, which ensures the unbiasedness of the estimating equations associated with the APQL function.
	%     Conditional on $(S_{it}, A_{it}, \bm{\alpha}_{i\cdot})$, $Y_{it}$ has density
	%    \[ f(Y_{it}|\bm{\beta}, \bm{\alpha}_i, \phi) = \exp\left \{ (Y_{it}\eta_{it} -b(\eta_{it})/{\phi} + c(Y_{it}, \phi)\right \} \]
	%    and $b(\eta)$ is twice  continuously differentiable and there exists a positive constant $c_0$ such that $c_0^{-1} \leq b''(\eta) \leq c_0$.
	\item Denote $\ell_1(\bm{\beta},\bm{\alpha})= \sum_{i=1}^n\sum_{t=1}^{m_i}\ell(Y_{it}, S_{it}, A_{it}; \bm{\beta}, \bm{\alpha}_{i\cdot},\phi)$.  We need the following regularity conditions:
	\label{ap:regularity}
	\begin{enumerate}[label=(\roman*)]
		\item As $N:=\sum_{i=1}^n m_i\to\infty$, $\bm{\beta}_0$ satisfies \\
		$\phantom{bbbbb} \quad N^{-1} E\left \{ \big[\nabla_{\bm{\beta}}\ell_1(\bm{\beta}_0,\bm{\alpha}_0)\big]^\T\nabla_{\bm{\beta}}\ell_1(\bm{\beta}_0,\bm{\alpha}_0)\right \}=O(1)$;\\
		$\phantom{bbbbb} \sup_{(\bm{\beta}, \bm{\alpha}) \in \Omega} \| N^{-1} \nabla ^\T_{\bm{\beta}}\nabla_{\bm{\beta}} \ell_1 (\bm{\beta}, \bm{\alpha})- E \left \{ N^{-1} \nabla ^\T_{\bm{\beta}}\nabla_{\bm{\beta}} \ell_1 (\bm{\beta}, \bm{\alpha})|\bm{\alpha}_0 \right \} \|_{{\rm F}}=o_P(1)$ $P_{\bm{\alpha}_0}$-almost surely, where $\|\cdot\|_{{\rm F}}$ denotes the Frobenius norm;\\
		and $\bm{M}_{\bm{\beta}\bm{\beta}}\triangleq -\liminf_{N\to\infty} E\left \{ N^{-1} \nabla ^\T_{\bm{\beta}}\nabla_{\bm{\beta}} \ell_1 (\bm{\beta}_0, \bm{\alpha}_0)\right \}$ is positive definite with all eigenvalues greater than $\delta_0>0$.
		
		\item  $\sup_i m_i^{-1} E\left \{ \nabla_{\bm{\alpha}_{i\cdot}}\ell_1(\bm{\beta}_0,\bm{\alpha}_0)^\T
		\nabla_{\bm{\alpha}_{i\cdot}}\ell_1(\bm{\beta}_0,\bm{\alpha}_0)\right \}=O(1)$.
		\item For $i=1,\ldots, n$, as $m_i\to\infty$, $\bm{\alpha}_{0i\cdot}$ satisfies\\
		$\phantom{bbbbb} \sup_{(\bm{\beta}, \bm{\alpha}) \in \Omega} \|  m_i^{-1} \nabla ^\T_{\bm{\alpha}_{i\cdot}}\nabla_{\bm{\alpha}_{i\cdot}} \ell_1 (\bm{\beta}, \bm{\alpha})- E\left \{  m_i^{-1} \nabla ^\T_{\bm{\alpha}_{i\cdot}}\nabla_{\bm{\alpha}_{i\cdot}} \ell_1 (\bm{\beta}, \bm{\alpha})|\bm{\alpha}_0 \right \} \|_{{\rm F}}=o_P(1)$ $P_{\bm{\alpha}_0}$-almost surely;\\
		and $\bm{M}_{\bm{\alpha}_{i\cdot}\bm{\alpha}_{i\cdot}}\triangleq -\liminf_{m_i} E\left \{ m_i^{-1} \nabla ^\T_{\bm{\alpha}_{i\cdot}}\nabla_{\bm{\alpha}_{i\cdot}} \ell_1 (\bm{\beta}_0, \bm{\alpha}_0)\right \}$ is positive definite with all eigenvalues greater than $\delta_0>0$.
	\end{enumerate}
	
	\item  The weights satisfy  $\max_{\{l\in\{1,\ldots,q\}:\sigma^2_l>0\}} |w_l|=O_P(1)$, where $\sigma_l^2$ is the $l$-th diagonal element of $\bm\Sigma$, the variance-covariance matrix of $\bm{\alpha}_{0i\cdot}$. \label{ap:weight}
	\item The tuning parameter $\lambda$
	satisfies  $\lambda =o\left \{ n^{-1}(\sum_{i=1}^n m_i^{-1})^{1/2}\right\}$.  \label{ap:lambda}
	%	\item For any $a\in \mathcal{A}$, almost surely, $P(A_{it}=a|\bm{H}_{it},\bm{\alpha}_{0i\cdot} ) \geq c_2^{-1}$ for a positive constant $c_2$.
\end{enumerate}

\begin{theorem}
	\label{thm:para}
	Suppose Assumptions \ref{ap:treatment}-\ref{ap:lambda} hold. As $n, \min_i \{m_i\}  \to \infty$, $(\hat{\bm\beta},\hat{\bm\alpha})$ satisfies $||\hat{\bm{\beta}}_\lambda - \bm{\beta}_0||= O_p(N^{-1/2})$ and  $||\hat{\bm{\alpha}}_{\lambda i} - \bm{\alpha}_{0i\cdot}|| = O_P(m_i^{-1/2})$, $i = 1, \ldots, n$. 
	%For any continuous function $g$,
	%\[ \left \| \frac{1}{n} \sum_{i=1}^n g(\hat {\bm{\alpha}}_{i\cdot})  -Eg(\bm{\alpha}_{1\cdot}) \right \| =o_p(1). \]
\end{theorem}
\noindent\textbf{Remarks.} Condition \ref{ap:regularity} is similar to the regularity conditions required in maximum likelihood estimation. 
In particular, when the covariates $S_{it}$'s are exogenous, it is easy to verify that \ref{ap:regularity} holds under the regularity conditions used in GLMM. 
Interestingly, Condition \ref{ap:regularity} will hold under many situations when $S_{it}$'s are endogenous; and importantly, these
situations can be verified.  For illustration purposes, we verify this condition in the Appendix  in two quite common
scenarios: (i) 
when $Y_{it}$ is binary and the distribution of $S_{it}$ directly depends on the latent random effects $\bm{\alpha}_{0i\cdot}$;  
(ii) when $Y_{it}$ follows Gaussian distribution and $S_{it} = Y_{i,t-1}$.

Theorem \ref{thm:para}  characterizes the asymptotic behavior of every $\hat{\bm{\alpha}}_{i\cdot}$ under the condition that  $\min_i{m_i}\to\infty$. This condition, however, can be relaxed if we are only interested in the asymptotic behavior of %the average
 $\hat{\bm{\alpha}}_{i\cdot}$ on average.  Specifically, we only require that the proportion of $m_i$'s that do not go to infinity goes to zero. Without loss of generality, suppose $m_1\leq m_2\leq \ldots\leq m_n$.  Let $k_n$ be the index so that $m_{k_n}$ is bounded, and $m_{k_n+1}\to\infty$.

\begin{corollary}
	\label{cor:para}
	Suppose \ref{ap:regularity}(iii) holds for $i= k_n+1,\ldots,n$, and the remaining assumptions in \ref{ap:treatment}-\ref{ap:lambda} continue to hold.  As $n, \min_{i> k_n} \{m_i\}  \to \infty$, suppose $k_n/n\to 0$. Then, $||\hat{\bm{\beta}} - \bm{\beta}_0||= O_P(N^{-1/2})$  and
	$$\frac{1}{n} \sum_{i=1}^n \|\hat {\bm{\alpha}}_{i\cdot} -\bm{\alpha}_{0i\cdot}\|^2 =O_P\left(\frac{k_n}{n}+\frac{1}{n}\sum_{i=k_n+1}^nm_i^{-1}\right). $$
\end{corollary}

Next, we present the  properties of the estimated personalized policies $\hat \pi_i$.
Specifically we consider both the conditional 
%both conditionally and marginally. In the conditional case, we study the 
expected outcome under $\hat{\pi}_{i}$ at each time point $t$ given $S_{it}=s_t$, and the marginal 
%That is, $\hat{\pi}_i$ is used to assign treatment for individual $i$ at time $t$. Since the expectation is conditional on $S_{it}=s_t$, by the Markovian property, treatment assignments prior to time $t$ do not affect outcome conditional on $S_{it}=s_t$.
%In the marginal case, we study the marginal 
expected outcome  assuming  $\hat{\pi}_{i}$ is used to make decision for user $i$ from the beginning to time point $t$. The results are stated in the theorem below.

\begin{theorem} \label{thm:bound}
	Assume all conditions in Corollary \ref{cor:para} hold.  Suppose the inverse link function of the corresponding exponential family distribution, $g^{-1}(\cdot)$, is H\"{o}lder continuous. That is, for any $\eta_1,\eta_2$ in the domain, $ |g^{-1}(\eta_1) - g^{-1}(\eta_2)| \leq L |\eta_1 - \eta_2|^{\gamma}$,
	where $L$ is a positive constant and $0< \gamma \leq 1$.  Then for any $t\geq 1$, as $n, \min_{i> k_n}\{m_i\}\to\infty$,
	\begin{align}
	\frac{1}{n}\sum_{i=1}^n[E_{\pi_{0i}}(Y_{it}|S_{it}=s_t, \bm{\alpha}_{0i\cdot})- E_{\hat\pi}(Y_{it}|S_{it}=s_t, \bm{\alpha}_{0i\cdot})] = O_P\left(\left[\frac{k_n}{n}+\frac{1}{n}\sum_{i=k_n+1}^nm_i^{-1}\right]^{\gamma/2}\right).
	\label{eqn:conditional}
	\end{align}
	In addition, assume	
	\begin{align}
	P_{\bm{\alpha}_{0i\cdot}}\left[\left\{\bm{\alpha}_{0i\cdot}: P\left(\arg\max_{a\in\mathcal{A}}Q(S_{i,t-1}(\underline{\pi}_{0i,t-2}),a;\bm{\beta}_0,\bm{\alpha}_{0i\cdot}) \mbox{ is unique }\big|\bm{\alpha}_{0i\cdot}\right)=1\right\}\right]=1, \label{eqn: unique}
	\end{align}
	where $S_{i,t-1}(\underline{\pi}_{0i,t-2})$ is the potential outcome of  $S_{i,t-1}$ that would have been observed were $\pi_{0i}$ used to make decision up to time point $t-2$.
	Then we have, 
	\begin{align}
	\frac{1}{n}\sum_{i=1}^n\left|E_{\pi_{0i}}(Y_{it}|\bm{\alpha}_{0i\cdot})- E_{\hat\pi_i}(Y_{it}|\bm{\alpha}_{0i\cdot})\right| = o_P(1).
	\label{eqn:marginal}
	\end{align}
\end{theorem}
\noindent\textbf{Remarks}. 
\begin{enumerate}
	\item The personalized policy $\pi_{0i}$ is  optimal in the conditional sense, in that it yields the maximal expected outcome if treatment assignment $A_{it}$ is consistent with $\pi_{0i}$ given $S_{it}$. As such, Equation (\ref{eqn:conditional}) describes the
	conditional optimality of estimated policies $\hat{\pi}_i$'s given the current information. We note that  $\pi_{0i}$ may not 
	necessarily be  optimal in a marginal sense after integrating out $S_{it}$, because 
		the distribution of $S_{it}$ depends on previous treatment assignment. Therefore, Equation (\ref{eqn:marginal}) in the
		above theorem implies consistency rather than optimality.
%	the marginal result (\ref{eqn:marginal}) shows the average consistency of $\hat{\pi}_i$ instead of optimality.
		\item Condition (\ref{eqn: unique}) implies that the optimal decision at time $t-1$ is unique almost surely, given that ${\pi}_{0i}$ were used to make decision at previous time points. This assumption is not needed to show consistency when $t=1$. 
\end{enumerate}

%%%%%%%%%%%%%%%%%%%%%%%%%%%%%%%%%%%%%%%%%%%%%%%%%%%%%%%%%%%%%%%
\section{Simulation study}
\label{sec:simu}

\subsection{Setup}
In this section, we examine the estimation properties of the maximum penalized-pseudo-likelihood estimator
$(\hat{\bm{\beta}},\hat{\bm{\alpha}})$ in 
(\ref{eqn:apql}) and the performance of the personalized policy $\hat \pi_i$ using simulation.

In a simulated trial, each user would be followed for $m=10,20,30$ time points for training purposes, with 10 additional
subsequent testing time points.
At time point $t$, user $i$ would receive one of three possible actions with equal probability, that is,
the actions were generated randomly with probabilities $(1/3, 1/3, 1/3)$; the actions $A_{it}$'s were then 
coded using two dummy variables and were centered.
The covariate process
$S_{it} = (X_{it}, t)$ included a binary endogenous variable $X_{it}\in \{-1, 1\}$, which would depend on the previous outcome
 $Y_{i, t-1}$, the previous action $A_{i, t-1}$ and the random effects $\bm\alpha$. 
 Specifically, we set $P(X_{i1}=1)= \alpha_{0i0}$, and  %generated $X_{it}$ with
 $$P(X_{it}=1|A_{i, t-1}, S_{i, t-1},\bm\alpha_{0i\cdot}) =\mbox{expit}((-3Y_{i, t-1}+2X_{i, t-1} - A_{i, t-1})/10+\alpha_{0i4}-\alpha_{0i5}+\alpha_{0i6}-\alpha_{0i7}),$$
 for $t\geq 2$, where expit$(\cdot)$ is the expit function, $\alpha_{0i0}\sim U(0,1)$, and $\alpha_{0ij}$ is the $j$-th component of $\bm\alpha_{0i\cdot}$ for $j=1,\ldots,q$.
%and $\tilde t$ denotes a scaled time point by dividing $t$ with 20, to stabilize the estimation. 
%The endogenous covariate $X_{it}$ depends on the previous outcome $Y_{i(t-1)}$, previous action $A_{i(t-1)}$ and the true random effect parameter $\alpha$. To generate $X_{it} \in \{-1, 1\}$, $P(X_{it}=1|A_{i(t-1)}, H_{i(t-1)},\alpha) =expit(-0.3*Y_{i(t-1)}+0.2*X_{i(t-1)} -0.1*A_{i(t-1)}+\alpha_{i4}-\alpha_{i5}+\alpha_{i6}-\alpha_{i7})$. 
We considered both binary and continuous outcomes.
The conditional mean of the outcome was defined according to (\ref{eqn:mixed1}) where
$h_1(S_{it}, A_{it}) = h_2(S_{it}, A_{it}) = (1, S_{it}, A_{it}, S_{it}\otimes A_{it})$ and $\otimes$ denotes the Kronecker product, 
with logit and identity links respectively for the binary and continuous outcomes.  The continuous outcomes were generated
with an independent Gaussian noise with standard deviation 1.5.
%At each given time point for each user, one of the three possible actions will be randomly generated with equal probability.
%That is, the actions are generated randomly with equal probability $(1/3, 1/3, 1/3)$ and are centered to ensure robust estimates. The first action is regarded as the reference group, thus $A_{it}$ contains two dummy variables. 
The true fixed effects were specified by  $$\bm{\beta}_0 = (-1, 0.2, -1.5, 0.8, 0.7, 0.1, 0.2, -1.2, -1.4)^\T.$$
We considered two scenarios for the random effects $\bm{\alpha}_0$, which were generated from mean zero Gaussian:
we set variance-covariance matrix to be $diag(2, 0.1, 0.1, 3, 4, 4, 5, 10, 12)$ to represent a scenario with
non-sparse random effects, and  
$diag(2, 0.1, 0.1, 3, 0, 0, 5, 10, 12)$ with sparse random random effects.
We generated 200 simulated trials, each having $n=50$ users.  Once the random effects were sampled, they were treated as fixed parameters in the 50 users.

The estimation properties of the policy parameters based on the training data were evaluated using mean squared error (MSE), defined as
$\sum_{i=1}^n||\hat{\bm{\beta}}(\pi)+\hat{\bm{\alpha}}_{i\cdot}^\T(\pi)-(\bm{\beta}_{0}(\pi)+\bm{\alpha}_{0i\cdot}^\T(\pi))||_2^2/(n\times\mbox{dim}(\bm\beta_0(\pi)))$, where $\bm\beta(\pi)$ is the sub-vector of $\bm\beta$ involved in policy $\pi$ (i.e. coefficients of $A_{it}$ and $S_{it}\otimes A_{it}$).
The quality of  decisions at the testing time points by the estimated policies was evaluated in terms of the expected conditional outcome under 
$\hat{\bm{\pi}}=\{\hat\pi_i\}$ at each testing time point $t$:

%for a given $X_{it}=x_{it}$ and averaged over individuals, calculated as 

% and the conditional mean value are used to evaluate the performance of different methods. The mean square error of the estimated policy parameter is $MSE_{\pi}=||(\hat{\bm{\beta}}_{\pi}+\hat{\bm{\alpha}}_{\pi}-(\bm{\beta}_{\pi}+\bm{\alpha}_{\pi}))||_2^2$, where the parameter with the subscription $\pi$ denote the policy parameter. 

%For a given set of estimated policies $\hat{\bm{\pi}}$, the primary measure of performance is the expected conditional outcome under $\hat{\bm{\pi}}$ at each time point $t$, for a given $X_{it}=x_{it}$ and averaged over individuals, calculated as 

%\begin{align}
%\label{eq:value}
$$V^{\hat{\bm{\pi}}}(s_t) \triangleq  \frac{1}{n}\sum_{i=1}^n         E^{\hat{\pi}_i}(Y_{it}|S_{it}=s_t)=\frac{1}{n}\sum_{i=1}^n Q(S_{it}=s_{t}, A_{it}=\hat{\pi}_i(s_{t});\bm\beta_0,\bm{\alpha}_{0i\cdot}),
$$%\end{align}
$t= m+1, m+2, \ldots, m+10$.
To facilitate comparison across scenarios, we standardized the expected outcome against the optimal policy $\bm\pi_0=\{\pi_{0i}\}$ and the worst
policy $\bm\pi_{\rm worst}=\{\pi_{\rm worst,i}\}$ and obtained the value ratio (VR) for the estimated policy $\hat{\bm\pi}$:

%The test set contains 10 time points following the training time points. The value in \eqref{eq:value} is scaled to $[0,1]$ by using the conditional value ratio of $\hat{\bm{\pi}}$, defined as

%\begin{align}
$$
VR^{\hat{\bm{\pi}}}(s_t) =\frac{V^{\hat{\bm{\pi}}}(s_t) - V^{\bm{\pi}_{\rm worst}}(s_t)}{V^{\bm{\pi}_0}(s_t) - V^{\bm{\pi}_{\rm worst}}(s_t)}.
$$
%\end{align}
%where $\pi_0$ is the optimal policy and $\pi_{worst}$ is the worst policy. 
% A secondary performance measure is the proportion of times that the actions chosen by the estimated policies match the optimal set of policies (called action ratio (AR)).

\subsection{Comparison Methods}

In the simulation, we  considered some existing methods as alternatives to the proposed personalized policy learning method, which 
shall be denoted as \textsf{PPL} in the followings.

Under the GLMM framework, instead of using the proposed algorithm described in Section~\ref{ssec:policy}, we used the ``glmer'' function in the lme4 package in R \citep{bates2014fitting}.
  This method shall be denoted as \textsf{glmer}.
The function ``glmer'' would involve approximating the marginal likelihood by integrating over the random effects.  
This could be problematic in situations with a large number of
random effects (thus having a high-dimensional integrals) and endogenous covariates.
%ddition, when the time-varying covariates were endogenous, the estimated marginal likelihood could be biased.
% The `glmer' package optimizes a marginal likelihood that integrates over the random effect for non-linear mixed models. Estimating the marginal likelihood requires approximating a high dimensional non-linear integral. However, when the time varying variable is endogenous and depends on the true random effect, the estimated marginal likelihood will be highly biased from the true marginal likelihood, especially when the number of time points is small and there is no sufficient data for estimation. 

In addition, we considered the regularized penalized quasi-likelihood (\textsf{rPQL}) approach developed by \citet{hui2017joint} for exogenous covariates as yet another 
alternative to estimating $(\bm\beta, \bm\alpha)$ under the GLMM framework.  While
\textsf{rPQL} also imposed a group lasso penalty, our 
proposed algorithm took a  different computational approach:  First,
we adopted the novel trust region method to solve the optimization problem; second, we updated the weights $w_l$'s iteratively
whereas \textsf{rPQL} would keep the weights at their initial values throughout the computation.

While the methods above would prescribe personalized policies, 
we also considered  using  generalized estimating equations (\textsf{GEE}) to
estimate a population-level effect, and developed a non-personalized policy by choosing actions maximizing the estimated
population mean.
We used an independence working correlation structure, so as to avoid bias under linear models with
endogenous variables; see
\cite{boruvka2018assessing}.

Finally, we examined the performance of an ``N-of-1''  approach whereby each user's personalized policy was estimated
by fitting a generalized linear model to the user's own data only.  That is, there was no borrowing information from across
users in this method with multiple generalized linear model (\textsf{MGLM}).  We  anticipated that \textsf{MGLM} would have difficulties
when $m$ was small, especially with Bernoulli outcomes.

\subsection{Simulation Results}
%\subsection{Estimation Properties}

Table \ref{tb:scena} compares the MSE of the policy parameters in the simulation scenario with non-sparse random effects.
Overall, the proposed \textsf{PPL} has the smallest MSE when $m=20, 30$.
%, and was very competitive with a small $m$.
Its superior performance to the other two GLMM-based methods (\textsf{glmer} and \textsf{rPQL})  indicates the computational advantages of using the trust region algorithm with iterated weights.
These three methods, as expected, improve with large $m$, that is, having more data points.

The ``N-of-1'' \textsf{MGLM}
performs poorly with binary outcome and when $m=10$ with continuous outcome.  Even with a moderate-to-large 
$m=30$, the method remains inferior to the other methods.  This signifies the importance of borrowing information from across
users, even though our goal is to produce different policies for different users.

Interestingly, \textsf{GEE} has the smallest MSE  when $m=10$ and performs relatively well with the larger $m$'s.
While it is somewhat surprising at first glance, we note that by avoiding estimating the random effects 
($\bm\alpha$ is estimated with $\bm 0$), \textsf{GEE} will induce the least variability and hence the MSE.
It is illuminating that the method's MSE does not improve as $m$ increases, when bias becomes dominating in the bias-variance
tradeoff.

%summarizes the mean square error of the policy parameter in the scenario (a) non-sparse random effect, and Table \ref{tb:scenb} summarizes the one in the scenario (b) sparse random effect, with sample size $n$=50 and number of training time points $m$=10, 20, 30 for both binary and continuous outcomes. 

\begin{table}[htbp]
	\caption{\label{tb:scena} Estimation properties under scenario with non-sparse random effects (Average MSE (SD) over $200$ simulation trials).}
	%: mean squared error of policy parameters $MSE_{\pi}$ estimated from PPL, GLMM, GEE, MGLM and rPQL.}
	\begin{tabular*}{\textwidth}{@{}l@{\extracolsep{\fill}}c@{\extracolsep{\fill}}c@{\extracolsep{\fill}}c@{\extracolsep{\fill}}c@{\extracolsep{\fill}}c@{\extracolsep{\fill}}c@{\extracolsep{\fill}}}
		\hline
		& \multicolumn{3}{c}{Binary} & \multicolumn{3}{c}{Continuous}\\
		\hline
		Method   & $m$=10 & $m$=20 & $m$=30 & $m$=10 & $m$=20 & $m$=30 \\
		\hline
		\textsf{PPL}& 8.22(3.31)&5.41(0.61)&4.70(0.58)&8.99(3.67)&3.69(0.47)&2.37(0.22)\\
		\textsf{glmer}& 43.39(34.16)&9.65(2.72)&6.39(1.17)&13.94(4.79)&4.87(0.65)&3.03(0.38)\\
		\textsf{GEE}& 7.87(2.92)&6.08(0.57)&6.17(0.52)&8.38(2.83)&5.94(0.38)&5.74(0.21)\\
		\textsf{MGLM}& >1E10&>1E10&>1E10&272.35(93.00)&36.24(35.00)&8.10(4.33)\\
		\textsf{rPQL}&8.71(3.9)&5.87(0.73)&5.23(0.65)&7.73(2.32)&5.31(0.30)&4.44(0.26)\\
		\hline
	\end{tabular*}
\end{table}

%    MGLM& 4.91E+29(5.25E+30)&3.48E+42(4.92E+43)&1.04E+42(1.34E+43)&272.35(93)&36.24(35)&8.1(4.33)\\

%In scenario (a), for both binary and continuous outcome, PPL has the lowest MSE when the number of decision points is sufficiently large, and the MSE decreases when $m$ increases. MGLM has the largest MSE compared with other methods, since there is no enough individual level information to learn the personalized policy parameters. For the binary outcome, the MSE of GLMM is obviously higher than the ones from PPL/GEE/rPQL, due to the biased marginal likelihood approximation. The `glmer' package optimizes a marginal likelihood that integrates over the random effect for non-linear mixed models. Estimating the marginal likelihood requires approximating a high dimensional non-linear integral. However, when the time varying variable is endogenous and depends on the true random effect, the estimated marginal likelihood will be highly biased from the true marginal likelihood, especially when the number of time points is small and there is no sufficient data for estimation. On the other side, PPL optimizes the adjusted quasi-likelihood and treats the random effect as fixed, thus is relieved from the integral bias. In scenario (b), When the number of decision points is sufficiently large, PPL has the lowest MSE; the models with penalty perform generally better than the ones without penalty.

Table \ref{tb:scenb} compares the methods under the scenario with sparse random effects.  The relative performance of the methods
is similar to that in Table \ref{tb:scena}, although the bias induced by \textsf{GEE} becomes more apparent as the variability in the data
is smaller in this scenario.  In particular, \textsf{PPL} and \textsf{glmer} has substantially smaller MSE in this scenario than when 
random effects are not as sparse.

\begin{table}[ht]
	\caption{\label{tb:scenb} Estimation properties under scenario with sparse random effects. (Average MSE (SD) over $200$ simulation trials).}
	\begin{tabular*}{\textwidth}{@{}l@{\extracolsep{\fill}}c@{\extracolsep{\fill}}c@{\extracolsep{\fill}}c@{\extracolsep{\fill}}c@{\extracolsep{\fill}}c@{\extracolsep{\fill}}c@{\extracolsep{\fill}}}
		\hline
		& \multicolumn{3}{c}{Binary} & \multicolumn{3}{c}{Continuous}\\ \hline
		Method   & $m$=10 & $m$=20 & $m$=30 & $m$=10 & $m$=20 & $m$=30 \\
		\hline
		\textsf{PPL}& 7.75(3.57)&4.41(0.80)&3.69(0.69)&7.17(3.09)&2.44(0.46)&1.33(0.20)\\
		\textsf{glmer}& 44.56(38.85)&8.80(2.50)&5.73(1.50)&11.75(4.08)&3.48(0.64)&1.86(0.33)\\
		\textsf{GEE}& 7.24(3.06)&5.04(0.71)&5.11(0.64)&7.15(2.98)&4.85(0.37)&4.67(0.23)\\
		\textsf{MGLM}& >1E10& >1E10& >1E10&274.92(105.00)&34.73(31.90)&7.19(1.78)\\
		\textsf{rPQL}& 8.04(3.86)&4.89(0.92)&4.22(0.82)&6.31(1.96)&4.42(0.40)&3.08(0.32)\\
		\hline
	\end{tabular*}
\end{table}

%MGLM& 7.63E+28(1.04E+30)&4.16E+41(5.89E+42)&4.85E+42(5.77E+43)&274.92(105)&34.73(31.9)&7.19(1.78)\\

%\subsection{Decision Qualities}

\begin{figure}[H]
	\centering
	\includegraphics[width=\textwidth]{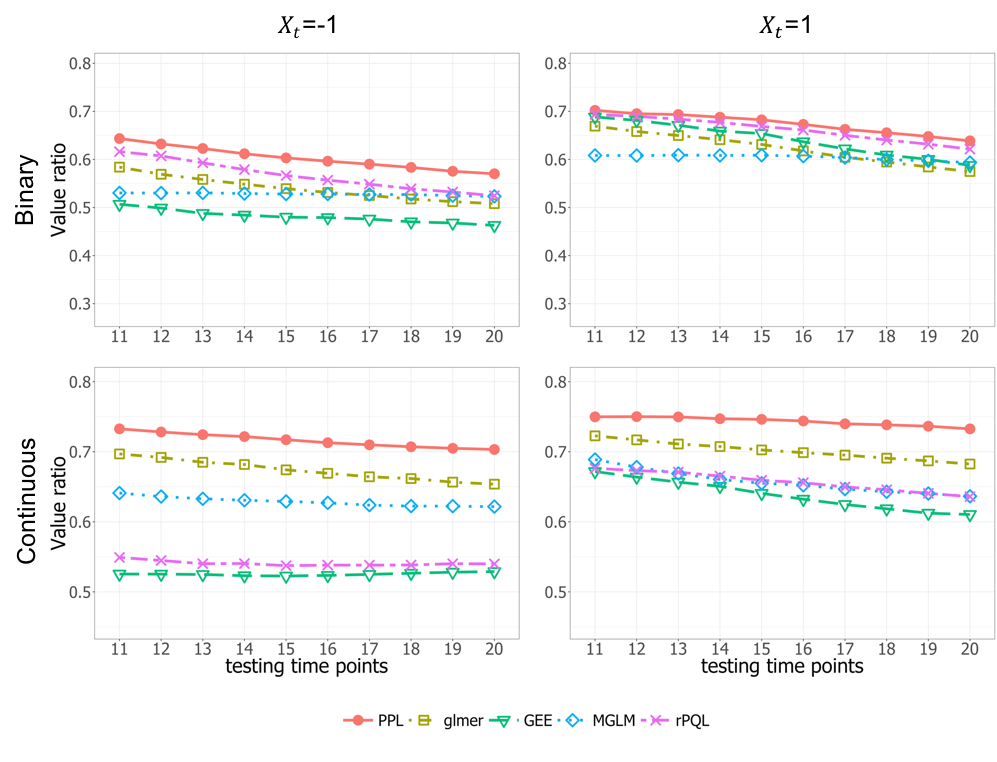}
	\caption{Value ratio at each testing time point in the simulation with $m=10$ under scenario with non-sparse random effects by different
	$X_{t}$.	}
%	: value ratio over test time points conditioning on different time varying variable $X_t$ values for binary outcome (up), continuous outcome (down). 
%\color{blue}{Can we make this figure smaller to save space?  Replace "GLMM" in the legend with "glmer".  Relabel the x-axis as "testing time points".  Relabel y-axis as "Value ratio".  The words "Continous",	"Binary", "X=-1", etc seem disproportionately big.  Can you also have $X_t$ instead of $X$? please rescale to make them look nice.  The vertical bars are confusing and not	informative - please remove.  Also, if possible, use darker lines (lwd=3) or darker colors - they don't show well on papers}
	\label{fig:vrnonsparse}
\end{figure}

\begin{figure}[H]
	\centering
	\includegraphics[width=\textwidth]{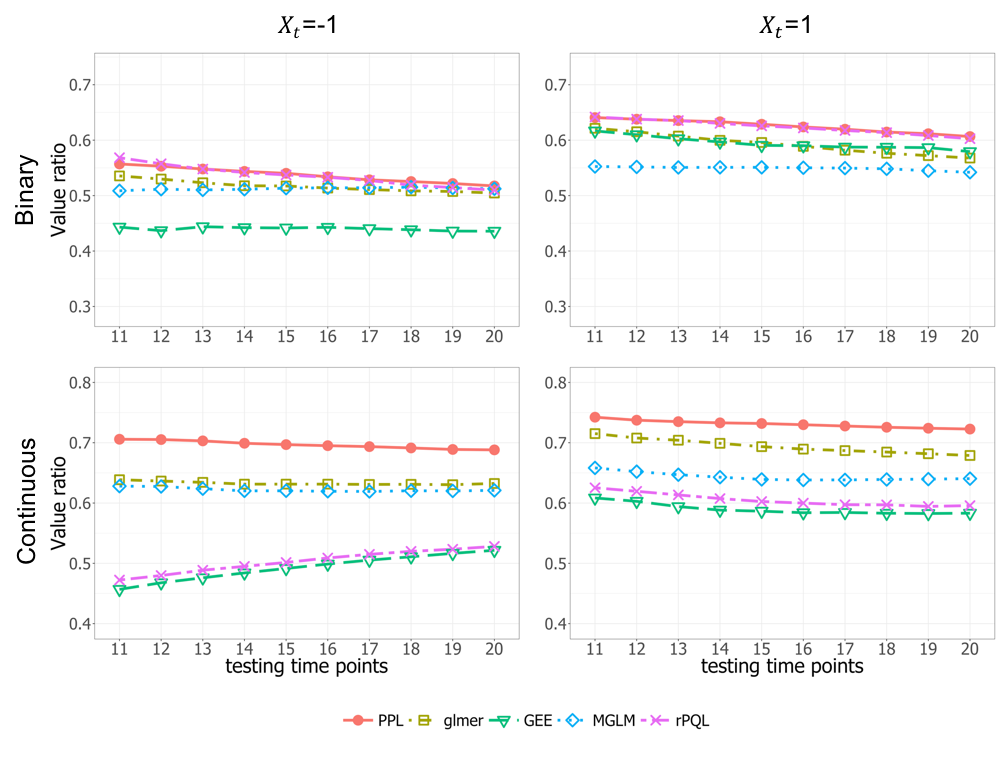}
	\caption{Value ratio at each testing time point in the simulation with $m=10$ under scenario with sparse random effects by different
	$X_{t}$.	}
%	: value ratio over test time points conditioning on different time varying variable $X_t$ values for binary outcome (up), continuous outcome (down). 
%\color{blue}{Can we make this figure smaller to save space?  Replace "GLMM" in the legend with "glmer".  Relabel the x-axis as "testing time points".  Relabel y-axis as "Value ratio".  The words "Continous",
%	"Binary", "X=-1", etc seem disproportionately big.  Can you also have $X_t$ instead of $X$? please rescale to make them look nice.  The vertical bars are confusing and not	informative - please remove.}}
	\label{fig:vrsparse}
\end{figure}

To compare the decision quality of the five methods, 
Figures \ref{fig:vrnonsparse} and \ref{fig:vrsparse} plot the simulated mean value ratio at the testing time points following $m=10$ training time points from each user, respectively under  non-sparse
random effects and sparse random effects.

The proposed \textsf{PPL} has the largest value ratio for each possible state $X_t$ for both binary and continuous outcomes.
That \textsf{GEE} producing the smallest MSE when $m=10$ does not translate into good decision quality, as
the method has the smallest value ratio uniformly in our simulation, when compared to all other personalized policy methods.
This serves as an important illustration how simply considering personalized policy,
 as opposed to personalized decisions (which \textsf{GEE} also prescribes),  could lead to
potentially radical gain.  It is interesting to note that methods that induce large variability in estimation can be quite
competitive; for example, \textsf{MGLM} and \textsf{glmer} for continuous outcome when $X_t=1$.  It is due to the fact that
the decision quality largely relies on correctly estimating the sign of the random effects, not the magnitude.  Therefore, one ought to
examine both the estimation properties and decision quality in the comparison of methods.  Overall, our simulation results indicate
the proposed \textsf{PPL} win in these terms.
%\color{blue}{Can we say that the relative performance of the methods is similar when $m=20,30$}
The relative performance of the methods is similar when $m=20,30$, and the results are presented in Section S3 of the Supplementary Material.

 %the mean and standard deviation of VR over test time points conditioning on different values of the time varying variable $X_t$ in (a) non-sparse random effect scenario, and Figure \ref{fig:vrsparse} illustrates the one in (b) sparse random effect scenario. The sample size $n$ is 50 and the number of decision points $m$ is 10 for both binary and continuous outcome. 
 
% If the data contains personalized effect, then estimating a one-fit-all policy to assign action is not ideal. For example, policies estimated from GEE obtain lowest VR compared to other methods account for personalized effect. Among the methods account for personalized effect, PPL achieves highest VR conditioning on the $t$ and $X_t$. When the outcome is binary, the rank of VR values: PPL>rPQL>GLMM>MGLM>GEE. Although GLMM and MGLM has a high estimation error, they capture the sign of personalized effect correctly, thus still achieve reasonable VR value. When the outcome is continuous, the performance of rPQL is unsatisfying, while PPL still achieves a higher VR value than GLMM. 

%The performance of all methods seems to be hampered when the true random effect is sparse. If the outcome is binary, then the models with the penalty term, such as PPL and rPQL perform better than GLMM. If the outcome is continuous, the models with the penalty term still perform better, among which PPL is better than rPQL.

\section{Application}
\label{sec:intell}

\begin{figure}[H]
	\begin{center}
		\includegraphics[width=0.8\textwidth]{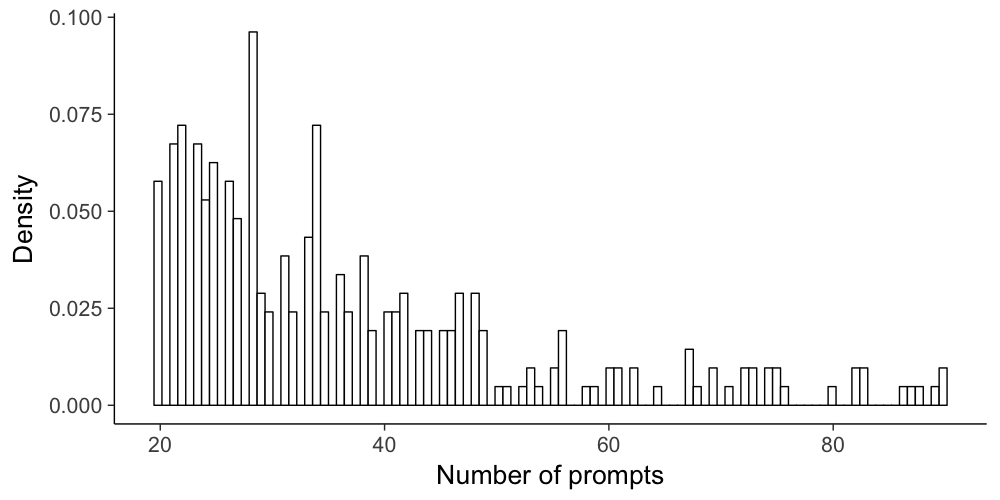}
	\end{center}
	\caption{The distribution of the number of prompts in 294 IntelliCare users. %\color{blue}{y-axis, relabel to "Density". x-axsis, relable "Number of prompts". Also make it smaller to save space}
	}
	\label{fig:promprs}
\end{figure}

We apply the proposed \textsf{PPL} to estimate the best personalized push schedule in 294  users,
who have received at least 20 prompts to complete the patient-health questionnaire since they downloaded the  Hub app.  %, that is, $m_i \geq 20$.  
Since the prompts were scheduled on  7-day intervals, this would represent a subsample of users with at least 20 weeks of app use.
The distribution of the number of prompts in these users is  shown in Figure \ref{fig:promprs}.  
In the data, we tracked the timestamp of when a prompt was sent.  For the purpose of this analysis, we grouped the time of prompt into
four periods: Night ($a_1$): from midnight to 6:00am; Morning ($a_2$): from 6:00am to noon; 
Afternoon ($a_3$): from noon to 6:00pm; Evening ($a_4$): from 6:00pm to midnight.
The observed proportions of the four periods were respectively 0.10, 0.23, 0.35, and 0.32.
Using $a_1$ as the reference group, we used three dummy variables, centered by the observed proportions, 
to code the actions $a_2, a_3$, and $a_4$ in model fitting.

%In this section, we apply the PPL method to the IntelliCare data. The goal is to develop a personalized policy for each individual that can be used to recommend the best time for sending out a prompt so as to enhance the response rate to the PHQ-4 questionnaire. IntelliCare is a suite of 12 mobile health applications and a centralized application named `Hub' for users self-evaluation and self-relief of depression and anxiety. User behavior and response patterns are collected using questionnaires designed specifically by domain experts. Prompts for filling the questionnaires are randomly sent to users. The first prompt is scheduled on the day after a user first launches the Hub app. If a user responds, then the next prompt will be a week later; otherwise, a follow-up prompt will be sent in the next 24 hours. The response rate over decision time points decreases over time. Therefore what time to send prompts that gain the highest response rate becomes an interesting research question. From the analysis of IntelliCare data, we observe heterogeneity of usage pattern across users. In addition, the response rate is different when prompts are sent at different hour of the day. This motivates our work to learn a personalized policy for each user for sending prompts to maximize current response rate at each prompt.

%In the data analysis we use data from 294 users who have 20 or more prompts, excluding those with greater than 100 prompts (0.087\%).  The outcome of interest $Y$ is the a binary variable indicating whether the user responses to a prompt or not. The action $A$ is a categorical variable with 4 levels indicating the hour intervals of sending prompt in a day, $[0, 6), [6, 12), [12, 18), [18, 24)$. So three dummy variables will be used to code the action in the model as $a_2$,$a_3$,$a_4$. The reference group is prompting during mid-night, that is the $[0,6)$ interval. The actions are centered. 
The state  $S_{it}$ at each time point consisted of three variables.  First, the number of times the Hub
was launched ({\em launches}) in the week prior to the prompt  was recorded.  Second, the timestamp indicated whether a prompt was sent on a weekday ({\em weekday}).
Third, the time point $t$ of the prompt was included as a predictor in the covariate process $S_{it}$.
With a binary response outcome, we estimated $(\bm \beta, \bm \alpha)$ under 
 model (\ref{eqn:mixed1})  with a logit link, $h_1(S,A)=(1, launches, t, weekday, A, launches \otimes A, weekday \otimes A, t \otimes A)$ and $h_2(S, A)=(1, A, launches \otimes A, weekday \otimes A, t \otimes A)$ using the first 80\% of the time points of each user as training data.
Since each user had at least 20 prompts, we had $m_i \geq 16$ in the training data for all 294 users.    

Table~\ref{tb:intell} summarizes the results of the model fit.
The positive fixed effects for $a_2, a_3, a_4$ suggest %are positive, suggesting 
prompts in the morning, afternoon, and evening tend to induce better response rate than those sent during the night 
(midnight to 6:00am).  The effects associated with these non-night periods are even greater on weekdays, indicated by
the positive (fixed) interaction between weekday and these periods.
While this result is not surprising, we also note substantial heterogeneity of the period effects and the $weekday$:period interactions, whose
SD$(\hat{\bm{\alpha}})$s have comparable magnitude to $\hat{\bm{\beta}}$.  This supports the needs for personalizing push schedule in our application.
% Therefore, it is advantageous to account for 

In contrast, 
for the $launches$:period interactions and the $t$:period interactions,
the fixed effects ($\hat{\bm{\beta}}$)   dominate
 the random effects; heterogeneity of the random effects coefficients  are measured by SD$(\hat{\bm{\alpha}})$.  
Based on the fixed effects,
the response rate decreases over time, by 0.20 in log-odds over $t=5$ time points. This is in line with findings in the literature; see \cite{helander2014} for example.
In addition,  every five additional {\em launches} of the Hub  in the prior week improves the log-odds of response to a night prompt by $1.52$.
Based on the negative coefficients of {\em launches}:period interactions, a large number of {\em launches} also seems to attenuate or even negate the effects of the time of prompts.  This suggests that for active users who engage the Hub often,
their response pattern is less sensitive to the time of the prompt.

%Prompting time, interaction of weekday and prompting time, interaction of launch and prompting time all highly contribute to the User variability. The variability is different when sending prompts at different time intervals of a day. For example, the variability is high if prompting in the morning compared with other prompting time.

\begin{table}[htbp]
	\caption{Model fit using the training data: $\hat{\bm{\beta}}$ is the coefficients of the fixed effects, and
		SD$(\hat{\bm{\alpha}})$ is the  standard deviation of the fitted individual random effects coefficients.  \label{tb:intell}
	}
	\centering
	\begin{tabular}{lcc}
		\hline
		\textbf{Variables} & $\hat{\bm{\beta}}$ &  SD$(\hat{\bm{\alpha}})$ \\
		%		$\hat{\bm{\beta}}_{GLMM}$ & $\hat{\bm{\beta}}_{rPQL}$ & $\hat{\bm{\beta}}_{GEE}$ \\
		\hline
		$Intercept$& -1.80&	1.31 \\
		$weekday$& 0.01 & --- \\
		$launches$ (per 5 times) &1.52& ---\\
		$t$ (per 5 time points)&-0.20& --- \\ 
		Morning ($a_2$) &1.65&	1.13 \\ 
		Afternoon ($a_3$)& 1.57&	0.95 \\
		Evening ($a_4$) &1.06&	0.78\\ 
		$weekday:a_2$& 0.73&	0.34\\
		$weekday:a_3$&0.16&	0.62\\
		$weekday:a_4$&0.66&	0.52\\
		$launches:a_2$&-2.46&	0.39 \\
		$launches:a_3$&-1.40&	0.21\\
		$launches:a_4$&-1.15&	0.43\\
		$t:a_2$&- 1.25&	0.66\\
		$t:a_3$&-0.96&	0.47\\
		$t:a_4$&-0.93&	0.44 \\
		\hline
	\end{tabular}
\end{table}

The quality of these personalized policies in the testing data is evaluated by the mean response rate under the policies estimated via inverse probability treatment weighted method averaged over all test time points.
The mean response rate according to \textsf{PPL} would have been 23\%, 
which compares favorably to other studies  
in light of the fact that all  testing points are at least 16 weeks from first download.
It has been reported that 
user engagement is in the range of 3\% to 15\% in the third month after download \citep{helander2014}. 
As a reference point, the  observed response rate in the testing data is 11\%.
In addition, we analyzed the prompt response data using the other methods with the same 80\%-20\% split of training and testing data, and obtained the mean response rate
14\%, 17\%, 14\%, and 8\% respectively for \textsf{glmer}, \textsf{GEE}, \textsf{MGLM}, and \textsf{rPQL}.

\section{Discussion}
\label{sec:diss}

This article makes several contributions.
 First, we have  shown personalized policies lead to higher value than non-personalized policy (i.e., \textsf{GEE}) in our simulation study, 
 and have clearly demonstrated substantial heterogeneity of the action effects in the prompt response data.  These results imply a paradigm shift and 
  call for the necessity of personalized policies,
which fundamentally differ from a single policy that may allow personalized decisions by tailoring.
Second, we propose a novel computational algorithm for the estimation of model parameters  under  GLMM  and for developing personalized policies.
 We have demonstrated, by simulation and in our data application,  that the algorithm leads to better estimation properties and decision quality when compared to
 some existing methods, namely \textsf{glmer} and \textsf{rPQL}.
Third, we have provided   theoretical justifications of the proposed \textsf{PPL} by examining its asymptotic properties under a fairly general set of assumptions.
In particular, we have established 
 consistency and optimality in the presence of endogenous covariate process, where the covariates may depend on previous outcomes, actions,  and even the latent
 random effects.  As endogeneity is  ubiquitous  in longitudinal mobile application usage (how many times a user launched the Hub app would likely depend on how he/she
had interacted with the Hub in the past), 
these theoretical results have broadened the applicability of  \textsf{PPL} to many practical situations.

\appendix
\section*{Appendix: Examples of endogenous covariates}
\label{app:example}

In this section, we verify condition \ref{ap:regularity} in two examples with endogenous covariates. In the first example, $Y_{it}$ is binary , and the distribution of $S_{it}$ directly depends on the random effects parameters $\bm\alpha_{0i\cdot}$. In the second example, $Y_{it}$ is Gaussian, and $S_{it} = Y_{i,t-1}$. For simplicity, we verify the condition with $n=1$ (since individuals are i.i.d.), and omit subscript $i$ from the notations.
In both examples, we consider a scalar mean zero random effects parameter $\alpha_0\in\mathbb{R}$, and the treatment $A_t\in\{-1,1\}$ is randomly assigned 
with $P(A_t=1)=P(A_t=-1)=1/2$ for $t\geq 1$.

{\bf Example 1.} For binary outcome $Y_t\in\{0,1\}$, suppose
$$g(E(Y_t|S_t, A_t)) = (\beta_0+\alpha_0)S_tA_t,$$
where $g(\cdot)$ is the logit link. Conditioning on $\alpha$, $S_t, t=1, \ldots, m$, are i.i.d. $N(\alpha_0, \tau^2)$.

Let $\ell(Y_t, S_t, A_t;\beta,\alpha)$ be the log-likelihood of Bermoulli distribution with mean $\frac{ e^{(\beta +\alpha)S_t A_t}}{1+ e^{(\beta +\alpha)S_t A_t}}$. Then $\ell_1(\beta,\alpha) = \sum_{t=1}^m\ell(Y_t, S_t, A_t;\beta,\alpha)$ satisfies
\begin{eqnarray*}
	\ell_1(\beta, \alpha)&=&\sum_{t=1}^m Y_t(\beta +\alpha)S_t A_t -\log \left \{ 1+e^{(\beta +\alpha)S_t A_t}\right \},  \\
	\nabla_{\beta} \ell_1(\beta, \alpha)&=& \nabla_{\alpha} \ell_1(\beta, \alpha)=\sum_{t=1}^m S_t A_t \left \{ Y_t- \frac{ e^{(\beta +\alpha)S_t A_t}}{1+ e^{(\beta +\alpha)S_t A_t}} \right \}, \\
	\mbox{and }	\nabla_{\beta}^\T \nabla_{\beta}\ell_1(\beta, \alpha)&=& \nabla_{\alpha}^\T \nabla_{\alpha} \ell_1(\beta, \alpha)= - \sum_{t=1}^m \frac{S_t^2e^{(\beta +\alpha)S_t A_t}}{\left \{ 1+e^{(\beta +\alpha)S_t A_t}\right \}^2}.
\end{eqnarray*}
It is easy to verify that
\[ E\left \{ \nabla_{\beta} \ell_1(\beta_0, \alpha_0)\right \}= E\left \{ \nabla_{\alpha} \ell_1(\beta_0, \alpha_0)\right \}=0, \]
\begin{eqnarray*} & & \frac{1}{m} E \left [ \left \{ \nabla_{\beta} \ell_1(\beta_0, \alpha_0)\right \}^2 \right ]= \frac{1}{m} E \left [ \left \{ \nabla_{\alpha} \ell_1(\beta_0, \alpha_0)\right \}^2 \right ]\\
	&=& \frac{1}{m} \sum_{t=1}^m E \left \{ \frac{S_t^2e^{(\beta_0 +\alpha_0)S_t A_t}}{\left \{ 1+e^{(\beta_0 +\alpha_0)S_t A_t}\right \}^2} \right \}= E \left \{ \frac{S_1^2e^{(\beta_0 +\alpha_0)S_1 A_1}}{\left \{ 1+e^{(\beta_0 +\alpha_0)S_1 A_1}\right \}^2} \right \}<\infty, \end{eqnarray*}
and \[ \bm{M}_{\beta \beta} =\bm{M}_{\alpha\alpha}= E \left \{ \frac{S_1^2e^{(\beta_0 +\alpha_0)S_1 A_1}}{\left \{ 1+e^{(\beta_0 +\alpha_0)S_1 A_1}\right \}^2} \right \} >0. \]
Finally, conditioning on $\alpha_0$, $\left \{ \frac{S_t^2e^{(\beta +\alpha)S_t A_t}}{\left \{ 1+e^{(\beta +\alpha)S_t A_t}\right \}^2}, t=1, \ldots, m \right \}$ are i.i.d., where $\alpha = \alpha_0 + u$. By the uniform law of large numbers theorem,
\[ \sup_{(\beta,\alpha)\in \Omega} \left | m^{-1} \nabla_{\beta}^\T \nabla_{\beta}\ell_1(\beta, \alpha)-E \left \{ \frac{S_1^2e^{(\beta +\alpha)S_1 A_1}}{\left \{ 1+e^{(\beta +\alpha)S_1 A_1}\right \}^2}\bigg  |\alpha_0\right \} \right |=o_P(1). \]
\\

{\bf Example 2.} Suppose
\[ Y_t |S_t, A_t, \beta_0,\alpha_0 \sim  N\{ (\beta_0 +\alpha_0) S_t A_t, \sigma^2\}, \]
where $S_t = Y_{t-1}$,  and $Y_0\equiv \mu_0$ is a constant. 
We consider $l(\cdot)$ to be the log-likelihood of Gaussian distribution. Below we show that condition \ref{ap:regularity} holds when 
\begin{eqnarray}
P_{\alpha_0} (|\beta_0 +\alpha_0| < 1)=1.  \label{eg2:c6}
\end{eqnarray}
Note that condition (\ref{eg2:c6}) is a sufficient condition for an AR(1) process to be stationary.

\begin{eqnarray*}
	\ell_1(\beta, \alpha)&=&-\frac{1}{2\sigma^2} \sum_{t=1}^m \left \{ Y_t-(\beta+\alpha) Y_{t-1}A_t \right \}^2, \\
	\nabla_{\beta} \ell_1(\beta, \alpha)&=& \nabla_{\alpha} \ell_1(\beta, \alpha)=\frac{1}{\sigma^2} \sum_{t=1}^m \left \{ Y_t-(\beta+\alpha) Y_{t-1}A_t \right \}Y_{t-1}A_t,\\
	\mbox{and }	\nabla_{\beta}^\T \nabla_{\beta}\ell_1(\beta, \alpha)&=& \nabla_{\alpha}^\T \nabla_{\alpha} \ell_1(\beta, \alpha)= - \frac{1}{\sigma^2}\sum_{t=1}^m Y_{t-1}^2.
\end{eqnarray*}
We can verify that
\[ E\left \{ \nabla_{\beta} \ell_1(\beta_0, \alpha_0)\right \}= E\left \{ \nabla_{\alpha} \ell_1(\beta_0, \alpha_0)\right \}=0. \]
Noticing that $E(Y_t^2)=\sum_{k=1}^{t-1} \sigma^2 E\left \{ (\beta_0+\alpha_0)^{2(k-1)}\right \}$, we have,
\begin{eqnarray*} & & \frac{1}{m} E \left [ \left \{ \nabla_{\beta} \ell_1(\beta_0, \alpha_0)\right \}^2 \right ]= \frac{1}{m} E \left [ \left \{ \nabla_{\alpha} \ell_1(\beta_0, \alpha_0)\right \}^2 \right ]\\
	&=& \frac{1}{m\sigma^2 } \sum_{t=1}^m E (Y_t^2) =  \frac{1}{m}\sum_{t=1}^m \sum_{k=1}^{t-1}E\left \{ (\beta+\alpha)^{2(k-1)}\right \}\\
	&=&\frac{1}{m}\sum_{t=1}^m (m-t+1) E\left \{ (\beta_0+\alpha_0)^{2(t-1)}\right \}
\end{eqnarray*}
which is $O(1)$ when (\ref{eg2:c6}) holds.

Since $ (m-t+1) E\left \{ (\beta_0+\alpha_0)^{2(t-1)}\right \}=m$ when $t=1$, we have
\[ \bm{M}_{\beta \beta} =\bm{M}_{\alpha\alpha}=\frac{1}{m}\sum_{t=1}^m (m-t+1) E\left \{ (\beta_0+\alpha_0)^{2(t-1)}\right \} \geq 1. \]
Finally, under condition (\ref{eg2:c6}), for any $\epsilon >0$,
\begin{eqnarray*}
	& & P \left \{ \sup_{(\beta,\alpha')\in \Omega} \left | \frac{1}{m} \sum_{t=1}^m Y_t^2 - E(Y_t^2|\alpha_0) \right | > \epsilon \bigg  |\alpha_0\right \}\\
	&= & P \left \{  \left | \frac{1}{m} \sum_{t=1}^m Y_t^2 - E(Y_t^2|\alpha_0) \right | > \epsilon \bigg  |\alpha_0\right \}\\
	& \leq & (\epsilon m)^{-2} E\left[ \sum_{t=1}^m \left \{ Y_t^2-E(Y_t^2|\alpha_0) \right \} \bigg  |\alpha_0 \right ]^2 \\
	&=& (\epsilon m)^{-2} 2\sigma^4 \sum_{t=1}^m \sum_{k=1}^{t-1} (\beta_0+\alpha_0)^{4(k-1)} + (\epsilon m)^{-2} 4\sigma^4 \sum_{t=1}^{m-1} \sum_{t'=t+1}^{m} (\beta_0+\alpha_0)^{2(t'-t)}\sum_{l=0}^{2t-1}(\beta_0+\alpha_0)^{2l} \\
	&=& (\epsilon m)^{-2} O(m) = o(1),
\end{eqnarray*}
which implies that
\[ \sup_{(\beta,\alpha)\in \Omega} \left | m^{-1} \nabla_{\beta}^\T \nabla_{\beta}\ell_1(\beta, \alpha)- E\left \{ m^{-1} \nabla_{\beta}^\T \nabla_{\beta}\ell_1(\beta, \alpha)|\alpha_0\right \} \right |=o_P(1). \]

% \bibliographystyle{jasanum2} 
% \bibliography{biblio}
%\bibliographystyle{plainnat}
%\bibliography{biblio}

\bigskip
\begin{center}
	{\large\bf SUPPLEMENTARY MATERIAL}
\end{center}
\begin{description}
\item Section S1 contains the estimation algorithm of policy parameters.  \\
\item Section S2 contains proofs of all technical results.  \\
\item Section S3 contains additional simulation results for decision quality comparison of the five methods when $m=20, 30$.
\end{description}

%\section{BibTeX}

%We hope you've chosen to use BibTeX!\ If you have, please feel free to use the package natbib with any bibliography style you're comfortable with. The .bst file agsm has been included here for your convenience. 

%\bibliographystyle{agsm}  

\bibliographystyle{apalike}  
\bibliography{biblio}
\end{document}